\documentclass[12pt,epsf]{article}
\usepackage{graphicx}
\begin{document}

\def\a{\alpha}
\def\b{\beta}
\def\c{\varepsilon}
\def\d{\delta}
\def\e{\epsilon}
\def\f{\phi}
\def\g{\gamma}
\def\h{\theta}
\def\k{\kappa}
\def\l{\lambda}
\def\m{\mu}
\def\n{\nu}
\def\p{\psi}
\def\q{\partial}
\def\r{\rho}
\def\s{\sigma}
\def\t{\tau}
\def\u{\upsilon}
\def\v{\varphi}
\def\w{\omega}
\def\x{\xi}
\def\y{\eta}
\def\z{\zeta}
\def\D{\Delta}
\def\G{\Gamma}
\def\H{\Theta}
\def\L{\Lambda}
\def\F{\Phi}
\def\P{\Psi}
\def\S{\Sigma}

\def\o{\over}
\def\beq{\begin{eqnarray}}
\def\eeq{\end{eqnarray}}
\newcommand{\gsim}{ \mathop{}_{\textstyle \sim}^{\textstyle >} }
\newcommand{\lsim}{ \mathop{}_{\textstyle \sim}^{\textstyle <} }
\newcommand{\vev}[1]{ \left\langle {#1} \right\rangle }
\newcommand{\bra}[1]{ \langle {#1} | }
\newcommand{\ket}[1]{ | {#1} \rangle }
\newcommand{\EV}{ {\rm eV} }
\newcommand{\KEV}{ {\rm keV} }
\newcommand{\MEV}{ {\rm MeV} }
\newcommand{\GEV}{ {\rm GeV} }
\newcommand{\TEV}{ {\rm TeV} }
\def\diag{\mathop{\rm diag}\nolimits}
\def\Spin{\mathop{\rm Spin}}
\def\SO{\mathop{\rm SO}}
\def\O{\mathop{\rm O}}
\def\SU{\mathop{\rm SU}}
\def\U{\mathop{\rm U}}
\def\Sp{\mathop{\rm Sp}}
\def\SL{\mathop{\rm SL}}
\def\tr{\mathop{\rm tr}}

\def\IJMP{Int.~J.~Mod.~Phys. }
\def\MPL{Mod.~Phys.~Lett. }
\def\NP{Nucl.~Phys. }
\def\PL{Phys.~Lett. }
\def\PR{Phys.~Rev. }
\def\PRL{Phys.~Rev.~Lett. }
\def\PTP{Prog.~Theor.~Phys. }
\def\ZP{Z.~Phys. }


\baselineskip 0.7cm

\begin{titlepage}

\begin{flushright}
IPMU-11-0074\\
UT-11-13\\
KEK-TH-1454\\
\end{flushright}

\vskip 1.35cm
\begin{center}
{\large \bf
    A Scalar Boson as a Messenger of New Physics
}
\vskip 1.2cm
Ryosuke Sato$^{1,2}$, Satoshi Shirai$^3$, and Tsutomu T. Yanagida$^{1,2}$
\vskip 0.4cm

{\it
$^1$Institute for the Physics and Mathematics of the Universe, University of Tokyo, \\
Kashiwa 277-8568, Japan\\
  
$^2$Department of Physics, University of Tokyo, \\ 
Tokyo 113-0033, Japan\\

$^3$ Institute of Particle and Nuclear Studies,\\
High Energy Accelerator Research Organization (KEK)\\
Tsukuba 305-0801, Japan\\
}

\vskip 1.5cm

\abstract{
Quantum corrections generate a quadratically divergent mass term for the Higgs boson in the Standard Model.
Thus, if the Higgs boson has a  mass of order 100 GeV, it implies the presence of a cut-off of the theory around TeV scale, and some particles associated with the new physics may appear around the cut-off scale $\L$.
However, if $\L$ is several TeV, it may be difficult to find such particles at the LHC.
In this paper, we consider a situation in which the new physics provides relatively light particles compared with the scale $\L$.
In such a situation, we show that diphoton event and four lepton event by the decay of the Higgs and/or a new particle have naturally large cross section,
and LHC may test the new physics in a considerably broad parameter region even if $\L$ is several TeV.
}
\end{center}
\end{titlepage}

\setcounter{page}{2}

\section{Introduction}

Higgs boson of mass of the order $100$ GeV is a crucial element in the standard model (SM).
However, its mass is not stable against quantum corrections. In fact, one-loop diagrams
induce quadratic divergences in the Higgs mass squared in the SM.
Thus, it is natural to consider some cut-off at around TeV scale, $\Lambda \simeq {\cal O}(1)$ TeV,
to keep the required ${\cal O}(100)$ GeV mass for the Higgs. If it is indeed the case,
we expect that LHC experiments find some particles associated with the new physics at the TeV scale.
However, the above argument does not necessarily predict the new physics within the reach of the LHC,
and in such unfortunate situation it may be very difficult to find new particles at LHC.

However, there are various new physics \cite{new} which predicts or requires relatively light particles
even if the cut-off scale $\Lambda$ is large as several TeV.
The purpose of this paper is to point out a possibility to test such new physics at LHC,
provided a relatively large cut-off scale of the new physics.

We consider, for simplicity, a gauge singlet scalar boson ${\phi}$ below 1 TeV generated 
by the new physics. However, the generalization to the other cases is straightforward. 
The low-energy effective theory is described by the SM $+$ one boson $\phi$. We take
mass of the scalar boson to be $m_\phi = 100~{\rm GeV}-1$ TeV and 
we consider the following possible interactions of the $\phi$,
\begin{equation}
{\cal L}_{\rm int} = {\kappa}\phi H^\dagger H + \sum_i \frac{\a_i}{\Lambda_i}\phi F_i^{\mu\nu} F_{i\mu\nu}.
\end{equation}
Here, $\a_i = g_i^2 /4\pi$ and $g_i$ is a corresponding gauge coupling constant.
The $\kappa$ in the first term is a dimension-one constant and we take $\kappa = {\cal O}(100)$ GeV.
The second term may arise from dynamics of the new physics.
This term gives large production cross section by gluon fusion.
Thus, we have three free parameters, $m_\phi, \kappa$ and $\Lambda_i$ in our effective approach.
And we show that LHC may test the new physics in a considerably broad parameter region even if
$\L_i$ is several TeV.

\section{Mass spectrum of scalar bosons and their decays}

Here, we consider the SM + one boson $\phi$ model.
For simplicity, we assume $\phi$ is a real scalar and a singlet under the SM gauge group.
We give the scalar potential as follows:
\beq
V = \frac{m_h^2}{2v^2} \left(H^\dagger H - \frac{v^2}{2}\right)^2 + \k\phi\left(H^\dagger H - \frac{v^2}{2}\right) + \frac{m_\phi^2}{2}\phi^2. \label{eq:potential}
\eeq
At the potential minimum, scalar fields have VEV with $H^\dagger H=v^2/2~(v=246~\GEV)$ and $\phi=0$.\footnote{
If we take the second term in Eq. (\ref{eq:potential}) as $- \k\phi\left(H^\dagger H - v'^2/2\right)~(v'\neq v)$, $\phi$ has non-zero VEV.
However, by an appropriate redefinition of the field and Higgs VEV $v$,
we can get the potential in Eq. (\ref{eq:potential}).
}
The first term of the potential is an ordinary Higgs potential.
The second term leads to the mixing and interaction between $H$ and $\phi$.
This term has an important role in the phenomenology of this model.
The last term is $\phi$ mass term.
This potential is bounded below if $|\k|< m_hm_\phi/v$ and $m_h^2 > 0$ is satisfied.

Furthermore, we introduce effective interaction terms between $\phi$ and the SM gauge bosons,
\beq
{\cal L}_{\rm int} = \frac{\a}{\L_{\g}}\phi F_{\m\n}F^{\m\n} + \frac{\a_s}{\L_{g}}\phi G_{\m\n}G^{\m\n}. \label{eq:interaction}
\eeq
Here, $F_{\m\n}$ and $G_{\m\n}$ are the field strength of photon and gluon, respectively.
In this paper, we do not specify sources of these effective interactions.
We can also write effective interaction term with $W$ or $Z$, such as $\phi Z_{\m\n}Z^{\m\n}$, $\phi Z_{\m\n}F^{\m\n}$ and $\phi W^\dagger_{\m\n}W^{\m\n}$.
In the following of this paper, we do not consider them for simplicity.

\subsubsection*{Mass spectrum of the model}

We decompose Higgs field into real scalar fields as
\beq
\left(
\begin{array}{c}
H_+\\
H_0
\end{array}
\right)
=
\frac{1}{\sqrt{2}}
e^{i\varphi^a T^a}
\left(
\begin{array}{c}
0\\
v+h
\end{array}
\right).
\eeq
Here, $\varphi^a~(a=1,2,3)$ are the Goldstone bosons which are eaten by $W$ and $Z$.
The mass and interaction terms of $\phi$ and $h$ are given by,
\beq
{\cal L} = -\frac{1}{2}m_\phi^2 \phi^2 + \k v \phi h - \frac{1}{2} m_h^2 h^2
+\frac{\k}{2}\phi h^2 + \frac{m_h^2}{2v} h^3 - \frac{m_h^2}{8v^2} h^4.
\eeq
We denote the mass eigenstate as ${\tilde \phi}$ and ${\tilde h}$.
They are defined by,
\beq
\left(
\begin{array}{c}
\tilde \phi \\
\tilde h
\end{array}
\right)
=
\left(
\begin{array}{cc}
\cos\theta & -\sin\theta \\
\sin\theta & \cos\theta \\
\end{array}
\right)
\left(
\begin{array}{c}
\phi \\
h
\end{array}
\right).
\eeq
The mass eigenvalues and mixing angle are given by,
\beq
m^2_{{\tilde\phi},{\tilde h}} = \frac{1}{2} \left( m_\phi^2 + m_h^2 \pm\sqrt{ (m_\phi^2 - m_h^2)^2 + 4\k^2 v^2 } \right),\\
\tan\theta = \frac{1}{2\k v}\left( \sqrt{ (m_\phi^2 - m_h^2)^2 + 4\k^2 v^2 } -m_\phi^2 + m_h^2\right).
\eeq
In the following of this paper, we assume $m_h^2 < m_\phi^2$.
In this case, $-\pi/2 \leq \theta \leq \pi/2 $.
When $|\k v| \ll |m_\phi^2-m_h^2|$, $\tan\theta \simeq \k v/(m_\phi^2-m_h^2)$.

\subsubsection*{Decay of $\tilde\phi$ and $\tilde h$}
Decay modes of $\tilde\phi$ and $\tilde h$ look like the SM Higgs because of its $h$ component.
The decay widths of $\tilde\phi$ and $\tilde h$ into SM fermions are given as follows :
\beq
\G({\tilde \phi} \to f{\bar f}) &=& \frac{N_f m_{\tilde\phi} }{8\pi} \frac{m_f^2}{v^2} \sin^2\theta \left( 1-\frac{4m_f^2}{m_{\tilde\phi}^2} \right)^{3/2}, \\
\G({\tilde h} \to f{\bar f}) &=& \frac{N_f m_{\tilde h}}{8\pi} \frac{m_f^2}{v^2} \cos^2\theta \left( 1-\frac{4m_f^2}{m_{\tilde\phi}^2} \right)^{3/2}.
\eeq
Here, $N_f=1$ for a lepton and $N_f=3$ for a quark.
If we assume $m_{\tilde\phi} > 2m_Z, 2m_{\tilde h}$, $\tilde\phi$ can decay to two massive gauge bosons or two $\tilde h$' s. The decay widths are given as follows :
\beq
\G(\tilde \phi \to W^+W^-) &=& \frac{\k^2}{16\pi m_{\tilde \phi}} \left(\frac{m_{\tilde \phi}^2 \sin\theta}{\k v}\right)^2 \left( 1-\frac{4m_W^2}{m_{\tilde\phi}^2}+\frac{12m_W^4}{m_{\tilde\phi}^4} \right) \sqrt{ 1-\frac{4m_W^2}{m_{\tilde \phi}^2} },\\
\G(\tilde \phi \to ZZ) &=& \frac{\k^2}{32\pi m_{\tilde \phi}} \left(\frac{m_{\tilde \phi}^2 \sin\theta}{\k v}\right)^2 \left( 1-\frac{4m_Z^2}{m_{\tilde\phi}^2}+\frac{12m_Z^4}{m_{\tilde\phi}^4} \right) \sqrt{ 1-\frac{4m_Z^2}{m_{\tilde \phi}^2} },\\
\G(\tilde \phi \to {\tilde h}{\tilde h}) &=& \frac{\k^2}{32\pi m_{\tilde\phi}} \left(\cos^3\theta-2\sin^2\theta \cos\theta + \frac{3m_h^2}{\k v}\sin\theta\cos^2\theta \right)^2 \sqrt{ 1-\frac{4m_{\tilde h}^2}{m_{\tilde \phi}^2} }.
\eeq
$\G_{WW} : \G_{ZZ} : \G_{{\tilde h}{\tilde h}} \simeq 2:1:1$ is derived in the limit $m_{\tilde\phi} \gg m_{\tilde h},~m_Z$.
This is the result of the Goldstone boson equivalence theorem \cite{Cornwall:1974km}.

\begin{figure}[tbp]
\begin{center}
  \includegraphics[width=.5\linewidth]{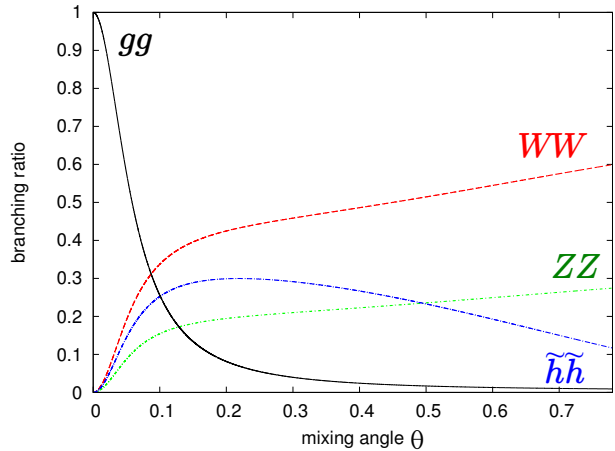}
  \end{center}
\caption{
The branching ratio of main decay process is shown.
In this figure, we set $\L_{{\tilde\phi},g}=1~\TEV$ and $\L_{{\tilde\phi},\g}^{-1}=0$, and fix mass eigenvalues with $m_{\tilde\phi} = 350~\GEV$ and $m_{\tilde h} = 115~\GEV$.
\label{fig:branch}}
\end{figure}

The decay processes $\tilde\phi$ and $\tilde h$ to $gg$ or $\g\g$ have contributions from new physics and SM loop effect.
By using $\L$ parameters defined in the Appendix, we can write the decay width of the process scalar particle to two photons or gluons simply as,
\beq
\G({\tilde \phi} \to gg) ~=~ \frac{2\a_s^2}{\pi} \frac{m_{\tilde \phi}^3}{|\L_{\tilde\phi,g}|^2},~~~~
\G({\tilde \phi} \to \g\g) ~=~ \frac{\a^2}{4\pi} \frac{m_{\tilde \phi}^3}{|\L_{\tilde\phi,\g}|^2},\nonumber\\
\G({\tilde h} \to gg) ~=~ \frac{2\a_s^2}{\pi} \frac{m_{\tilde \phi}^3}{|\L_{\tilde h,g}|^2},~~~~
\G({\tilde h} \to \g\g) ~=~ \frac{\a^2}{4\pi} \frac{m_{\tilde \phi}^3}{|\L_{\tilde h,\g}|^2}. \label{eq:effective}
\eeq
In Fig. \ref{fig:branch}, we show the branching ratio of the decay $\tilde\phi$.

\section{Constraints and signals of the model}
Here, we consider constraints and signals of the present model at hadron colliders, that is, the Tevatron and the LHC.
As we denoted in the previous section, $\tilde\phi$ and $\tilde h$ have similar coupling and decay mode to the SM Higgs.
Therefore, the SM Higgs search gives constraints on the model.
At a hadron collider, the SM Higgs is produced by gluon fusion (GF) and vector boson fusion (VBF) dominantly.
However, in the present model, VBF cross section is suppressed because of mixing angle $\theta$.\footnote{Here, we assume a coupling between the Higgs and transverse component of weak gauge boson is not enhanced by the operator $\phi W_{\m\n}W^{\m\n}$.
This assumption can be justified when $\L$ is a few TeV.
}
On the other hand, GF is enhanced because of a coupling between scalar particles and gluon.
Then, we discuss the collider signal for the Higgs produced by GF.

The $\tilde\phi$ production diagram by GF includes a vertex which gives a $\tilde\phi\to gg$ diagram.
Therefore, the production cross section is proportional to $\G({\tilde \phi}\to gg)$.
By using the narrow width approximation, we get the $\tilde\phi$ cross section as \cite{Georgi:1977gs},
\beq
\s(pp\to pp{\tilde \phi}) = \frac{\pi^2}{8m_{\tilde\phi}s}\G({\tilde\phi}\to gg) \int_0^1 dx_1 \int_0^1dx_2 ~\delta\left(x_1x_2-m_{\tilde \phi}^2/s \right) g(x_1)g(x_2).~\label{eq:xsecformula}
\eeq
Here, $\sqrt s$ is the center of mass energy, and $g(x)$ is the gluon distribution function of the (anti-)proton.
$\tilde h$ production cross section also obeys to similar formula.

\begin{figure}[tbp]
\begin{center}
  \begin{minipage}{.50\linewidth}
  \includegraphics[width=\linewidth]{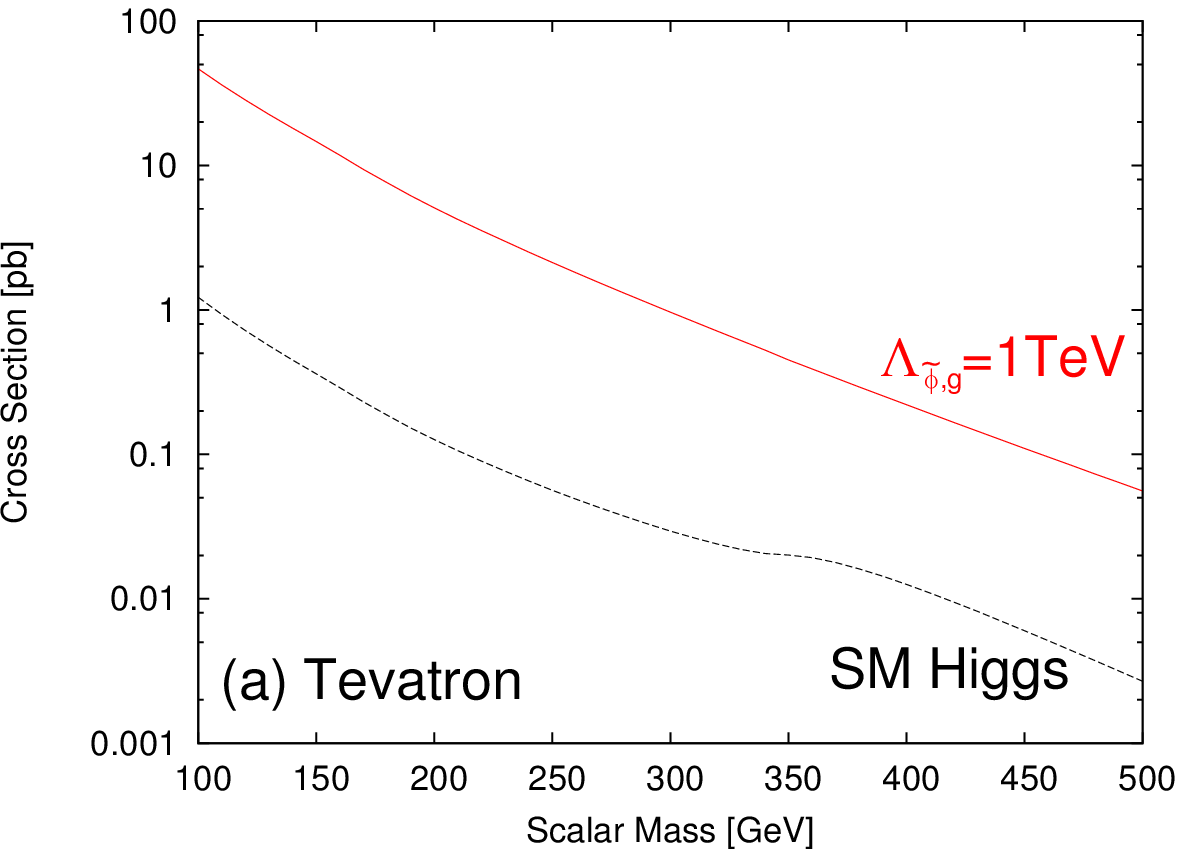}
  \end{minipage}
\\
\vspace{.5cm}
  \begin{minipage}{.50\linewidth}
  \includegraphics[width=\linewidth]{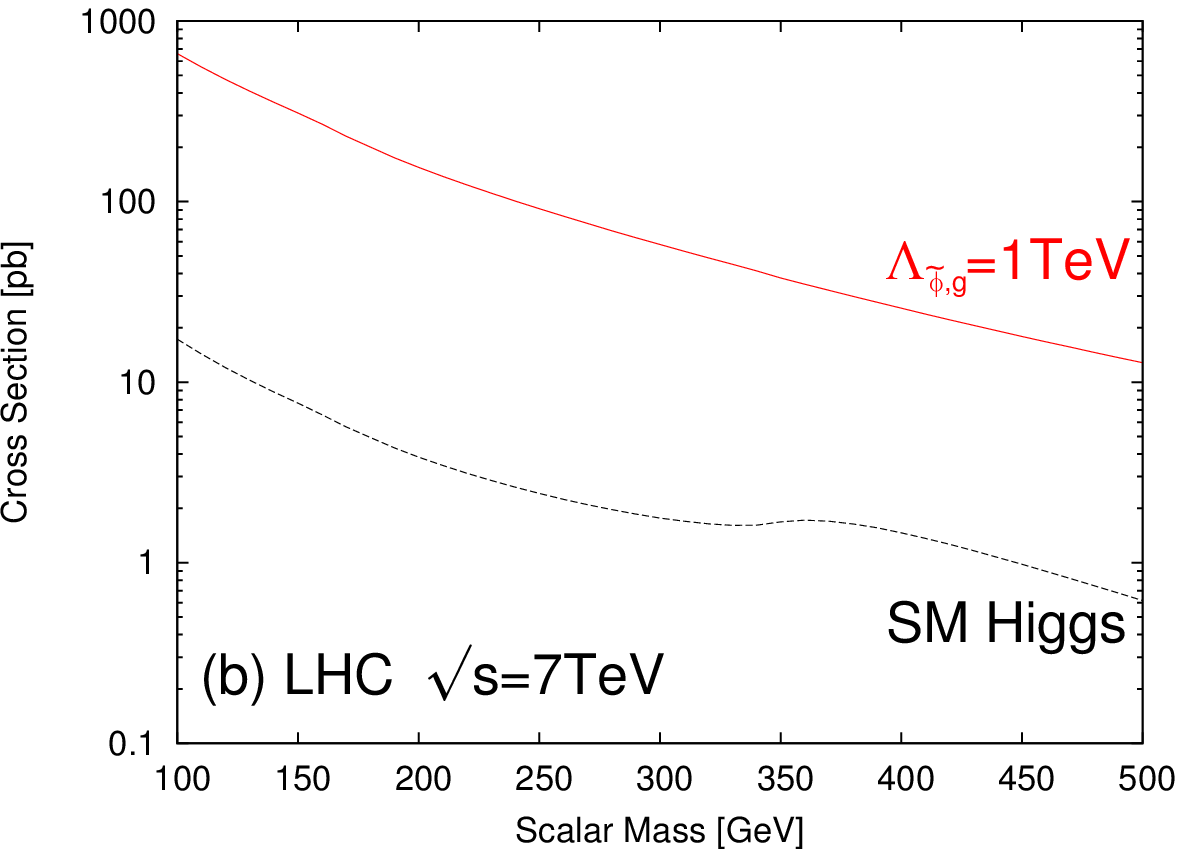}
  \end{minipage}
\\
\vspace{.5cm}
  \begin{minipage}{.50\linewidth}
  \includegraphics[width=\linewidth]{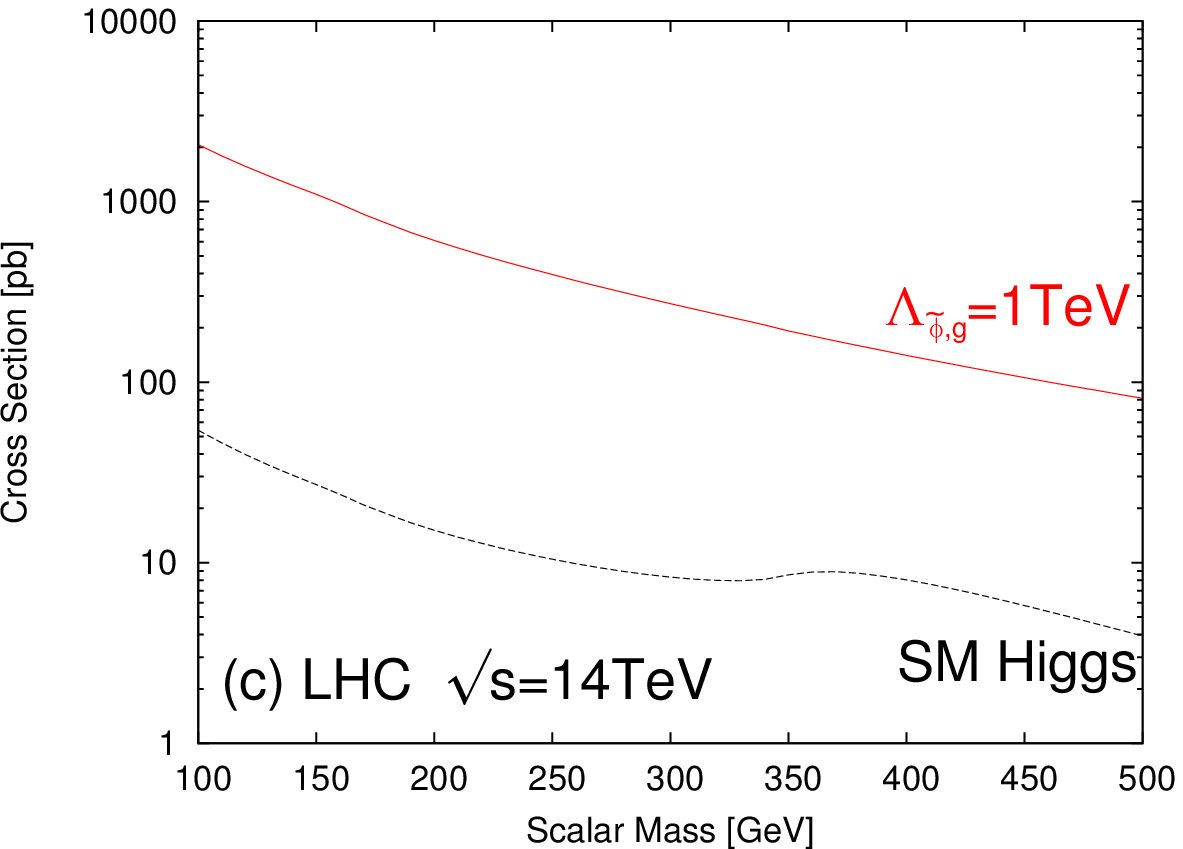}
  \end{minipage}
\end{center}
\caption{
$\tilde\phi$ ($\tilde h$) production cross section is shown as a function of $m_{\tilde\phi}$ ($m_{\tilde h}$).
In these figures, we set $|\L_{\tilde\phi,g}|=1~\TEV$ ($|\L_{\tilde h,g}|=1~\TEV$) at the Tevatron (a), the LHC $\sqrt{s} = 7~\TEV$ (b) and $\sqrt{s} = 14~\TEV$ (c).
The cross section is proportional to $|\L_{\tilde\phi, g}|^{-2}$ ($|\L_{\tilde h, g}|^{-2}$).
For reference, the SM Higgs production cross section in gluon fusion is also plotted in these figures.
\label{fig:xsec}
}
\end{figure}

In Fig. \ref{fig:xsec}, we show $\tilde\phi$ (or $\tilde h$) production cross section at the Tevatron and the LHC.
We estimated $\tilde\phi$ cross section as $\s(gg\to h_{\rm SM}) \times \G(\tilde\phi\to gg)/\G(h_{\rm SM}\to gg)$ by using Eq. (\ref{eq:xsecformula}).
The SM Higgs production cross section is calculated by the program HIGLU \cite{Spira:1995mt}.
A decay width without $\g\g$ and $gg$ mode of the SM Higgs is calculated by the program HDECAY \cite{Djouadi:1997yw}.
$\G(\tilde\phi\to gg)$ and $\G(\tilde\phi\to \g\g)$ are calculated at leading order.
Let us note this calculation is a rough estimate.
To argue precisely, we must specify high energy physics, and may have to calculate at next leading order.

\begin{figure}[tbp]
\begin{center}
  \includegraphics[width=0.5\linewidth]{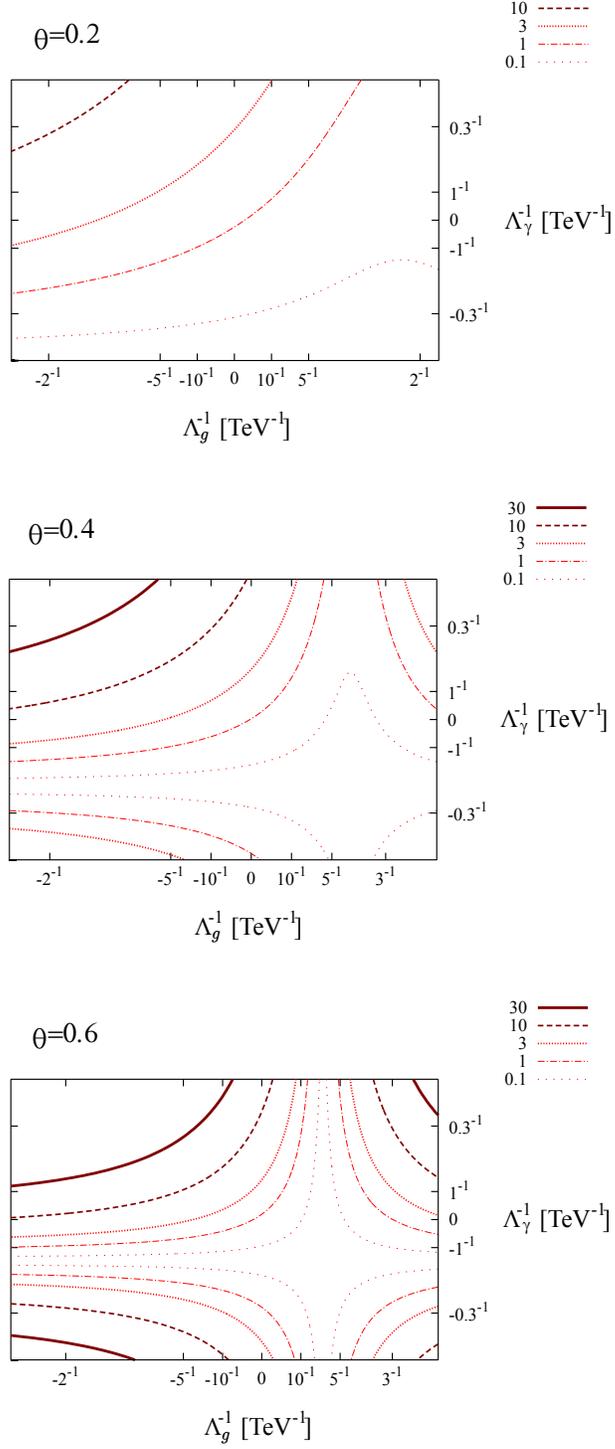}
\end{center}
\caption{
In each figure, we show $\sigma(gg\to\tilde h\to\gamma\gamma) / \sigma(gg\to h\to\gamma\gamma)|_{\rm SM}$.
We set $m_{\tilde h}=115$ GeV, $m_{\tilde\phi}=350$ GeV and $\theta$ = 0.2, 0.4 and 0.6.
This region avoids the constraints on other channels from LEP, Tevatron and ATLAS.
\label{fig:twophoton}
}
\end{figure}

First, let us consider constraints on the production of $\tilde h$.
The most stringent constraint comes from two photon search because of an enhancement of a branching ratio to two photons due to the operator $\phi F_{\m\n} F^{\m\n}$. In Fig. \ref{fig:twophoton}, we show the contour plot of $\left[ \s(gg\to{\tilde h}) \times {\rm Br}(\tilde h\to\g\g) \right] / \left[ \s(gg\to h_{\rm SM}) \times {\rm Br}(h_{\rm SM} \to\g\g) \right]$,
which should lower than 25~\cite{Peters:2010gd} at the Tevatron when $m_{\tilde h} = 115~\GEV$.
$\t\t$ channel are constrained for $\s \times {\rm Br} \lsim$ 6 pb~\cite{Benjamin:2010xb} at the Tevatron.
$\tilde h$ can be produced by a decay of $\tilde\phi$.
However, $\tilde h$ is sufficiently lighter than $\tilde\phi$,
therefore, unless mixing effect cancel out the SM loop and new physics contribution,
direct production of $\tilde h$ is a dominant process.

\begin{figure}[tbp]
\begin{center}
  \includegraphics[width=\linewidth]{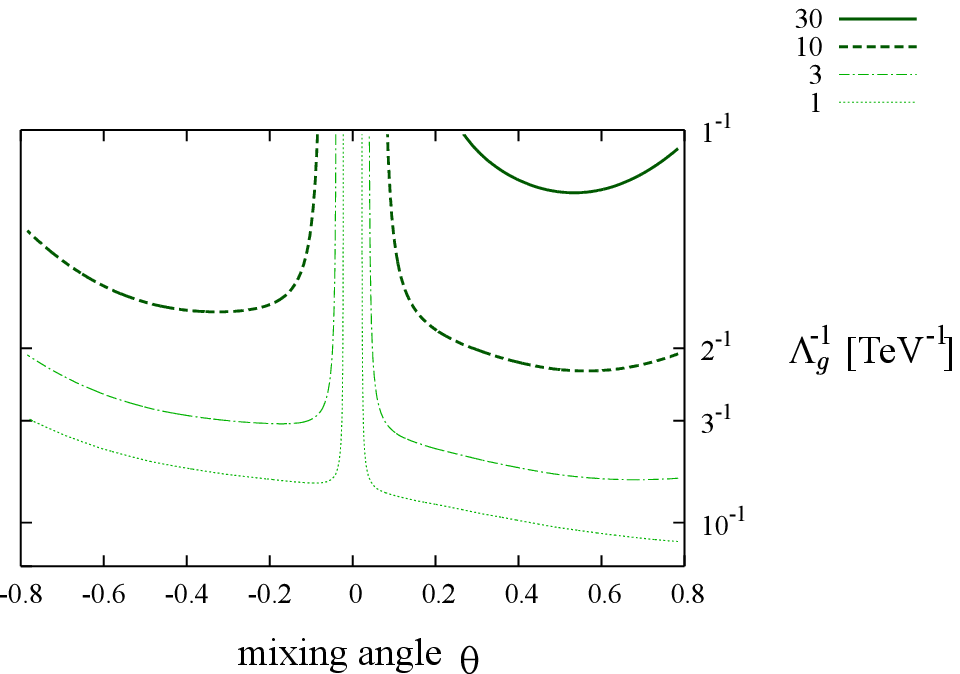}
  \end{center}
\caption{
We show $\sigma(gg\to{\tilde\phi} \to ZZ) / \sigma(gg\to h_{\rm SM} \to ZZ)|_{\rm SM}$
In this figure, we set $\Lambda_\gamma^{-1} = 0$.
The cross section is almost independent on $\Lambda_\gamma$.\label{fig:phitozz}
}
\end{figure}

Next, let us consider constraints on the production of $\tilde\phi$.
$\tilde\phi$ decays into two $W$'s or two $Z$'s mainly.
These channels give severe constraint.
In Fig. \ref{fig:phitozz}, we show the contour plot of $\left[\s(gg\to{\tilde\phi}) \times {\rm Br}(\tilde\phi\to ZZ)\right] / \left[\s(gg\to h_{\rm SM}) \times {\rm Br}(h_{\rm SM}\to ZZ)\right]$.
$WW$ channel is constrained on the upper bound 1.4 pb~\cite{Aaltonen:2010ws} at the Tevatron when $m_{\tilde\phi} = 350~\GEV$.
For this channel, ATLAS constrained the upperbound 14 pb~\cite{atlasWW} on the process $gg\to h_{\rm SM}\to WW$ at the LHC $\sqrt{s}=7~\TEV$.
$\tilde\phi \to \g\g$ is enhanced due to $\phi F_{\m\n} F^{\m\n}$.
This channel is constrained on the upper bound 0.007 pb at the Tevatron \cite{Aaltonen:2010cf}.
$gg$ channel also gives constraint, but this constraint is weak compared to $WW$ channel.

\section{Conclusion and discussion}

In this paper, we consider the situation in which the scale of the new
physics is several
TeV. Naively, it is difficult to find such a new physics at the LHC.
However, if relatively
light particles are given from the new physics, it is expected such a
particle provides a
possibility to test new physics at the LHC.

In the present model, the new particle and the standard Higgs boson are
mixed with each other.
By this mixing, the production and decay-mode are drastically changed,
compared to the standard Higgs model.
In some parameter region the production cross section can be ${\cal O}(10)$
times
larger than the standard model Higgs for $m_{\tilde h}$=115 GeV ($m_{\tilde\phi} = $ 350 GeV)
without conflict with the current experimental limits.

For the lighter $\tilde\phi,\tilde h$ which cannot decay into two $Z$ bosons,
enhancement of the branching fraction to two photons is very important for collider signal.
By the enhancement of the cross section and/or branching fraction to
photons,
the $\sigma\times {\rm Br}$ reach some ten times larger than the standard Higgs boson in
some parameter region and
it is possible to discover the light Higgs boson with the diphoton
channel even in gluon fusion process.
On the other hand, the vector boson fusion process is suppressed by the
mixing angle $\cos^2\theta$,
compared to the standard model Higgs.
Therefore by using information on forward-jet, this model can be
tested.

As for the heavier $\tilde\phi, \tilde h$,
the cleanest signals of such particles are four-leptons and/or diphoton
events.
The cross section times branching fraction $\sigma \times {\rm Br}(\tilde \phi \to ZZ)$ can be
some ten times larger than the standard model Higgs case.
In such a case, $\tilde\phi$ plays a very important role for $ZZ$ search such as
\cite{Abazov:2011td, CDF}.

In the present model, there are two particles $\tilde h, \tilde\phi$.
Therefore in some parameter region, both two can make clean signal,
e.g.,
115 GeV diphoton mass and 350 GeV four leptons mass.
In such a case, observation $\tilde\phi \to \tilde h\tilde h$ or $\tilde h \to \tilde\phi \tilde\phi$ is the
most crucial test for the present model.
\\
\\
{\bf Note added:} After the completion of this work we received a paper \cite{Fox:2011qc},
which has some overlap with our paper.

\section*{Acknowledgements}
This work  was supported by the World Premier 
International Research Center Initiative (WPI Initiative), MEXT, Japan.
The work of SS and RS is supported in part by JSPS Research
Fellowships for Young Scientists.

\appendix
\section{Relationships among $\L$'s}

Here, we denote relationships among $\L$'s in Eqs. (\ref{eq:interaction}) and (\ref{eq:effective}).
\beq
\frac{1}{\L_{\tilde\phi,g}} &=& \frac{\cos\theta}{\L_{g}} + \sin\theta \sqrt{\frac{G_F}{512\sqrt{2}} }~A_{1/2}\left( \frac{m_{\tilde\phi}^2}{4m_t^2} \right), \\
\frac{1}{\L_{\tilde\phi,\g}} &=& \frac{\cos\theta}{\L_{\g}} + \sin\theta \sqrt{\frac{G_F}{32\sqrt{2}}  }\left[ \frac{4}{3}A_{1/2}\left( \frac{m_{\tilde\phi}^2}{4m_t^2} \right) + A_1\left( \frac{m_{\tilde\phi}^2}{4m_W^2} \right) \right], \\
\frac{1}{\L_{\tilde h,g}} &=& \frac{-\sin\theta}{\L_{g}} + \cos\theta \sqrt{\frac{G_F}{512\sqrt{2}} }~A_{1/2}\left( \frac{m_{\tilde h}^2}{4m_t^2} \right), \\
\frac{1}{\L_{\tilde h,\g}} &=& \frac{-\sin\theta}{\L_{\g}} + \cos\theta \sqrt{\frac{G_F}{32\sqrt{2}}  }\left[ \frac{4}{3}A_{1/2}\left( \frac{m_{\tilde h}^2}{4m_t^2} \right) + A_1\left( \frac{m_{\tilde h}^2}{4m_W^2} \right) \right].
\eeq
Here, we use a notation in Ref.~\cite{Djouadi:2005gi}.
$A_{1/2}$ and $A_1$ are defined by, 
\beq
A_{1/2}(\tau) &=& \frac{2}{\t^2}(\t + (\t-1)f(\t) ), \\
A_1(\tau) &=& -\frac{1}{\t^2}(2\t^2 + 3\t + 3(2\t-1)f(\t) ).
\eeq
$A_{1/2}$ and $A_1$ stand for SM fermion and massive gauge boson loop effects, respectively.
$f(\tau)$ is given by,
\beq
f(\tau) = 
\left\{
\begin{array}{ll}
\arcsin^2\sqrt{\t} & (\tau\leq 1) \\
-\displaystyle\frac{1}{4}\left[ \log\left(\frac{1+\sqrt{1/\tau}}{1-\sqrt{1/\tau}}\right)-i\pi \right]^2 & (\tau>1)
\end{array}
\right..
\eeq

\end{document}